\documentclass{vospwks2007}
\usepackage{graphicx}

\title{Common Methods of Stellar Spectra Analysis  and their Support in VO}
\author{Petr \v{S}koda}
\affil{Astronomical Institute of the Academy of Sciences of the Czech
Republic, v.v.i. \\ Fri\v{c}ova 298, Ond\v{r}ejov, CZ-251\,65}

\begin{document}

\keywords{spectroscopy, stellar continuum, spectral lines, radial velocity, web services, SSAP , Virtual Observatory}
\maketitle

\begin{abstract} 

The proper scientific analysis of a large amount of stellar
spectra requires certain capabilities of the analysing tool (e.g. precise
semi-automatic fitting for normalisation of the continuum or line-list
assisted measurement of spectral lines) as well as  flexible list-driven
datafile handling.

While most astronomical legacy packages comprise powerful analysing and data
management features allowing rapid processing of quite complex data, current
VO tools are lacking support of  even basic capabilities commonly used in
stellar spectroscopy. 

Especially the high resolution optical spectroscopy of stars with rapid line
profile variations or of those with complicated emission profiles benefits from
a number of specific methods unsupported in todays Virtual Observatory (VO)
tools and even lacking definition in VO spectral protocols. 

In our contribution we  identify these techniques, describe their possible
implementations and finally give a short overview of several VO-compatible
tools emphasising their deficiencies and comparing with the capabilities of
common legacy packages.  

\end{abstract}
\section{Introduction} 

There is a wealth of calibrated astronomical spectra accessible in current
data archives, however most of them are not suitable for direct physical
analysis.  Being prereduced by some automatic pipeline or individually by
manual reduction, they  are mostly stored in archives in form of  FITS file
or ASCII table as the relation of intensity in arbitrary numbers
(instrumental counts, digital numbers etc.) and wavelength (or frequency).

The  scientific analysis of such  spectra requires further processing by the
variety of different methods. In certain studies a huge number of spectra has
to be collected from different servers (e.g. in different spectral regions )
and transformed into  common units. Often the transformed spectra or variables
measured in them (e.g. equivalent width) are visually inspected as the time
series ordered by time of observation or by the orbital phase.

The tools with built-in VO protocols may considerably increase the efficiency of
astronomical analysis by eliminating the data collection,
aggregation and  conversion and by allowing the astronomer to fully  concentrate on
the visual properties of data (e.g. searching the variability of line
profiles or constructing the radial velocity curve). To be accepted quickly
by the wide astronomical community a number of methods and recipes has to be
implemented in a manner very similar to  the  current legacy applications.
We attempted to identify the most useful methods and give  hints for their
implementation in VO clients.

\section{Common methods of astronomical spectroscopic analysis}

\subsection{Rectification} 

Usually the first step in data analysis is the rectification of the spectrum
(normalisation of the continuum to level 1.0). Although absolutely
flux-calibrated spectrum is common for space instruments, it is rather
exceptional in ground-based optical stellar spectroscopy (used mainly for
determination  of spectral type from low resolution spectra). 

Usually, simple polynomials of low degree are applied for this purpose
but there is a number of problems that have to be addressed:

\begin{description} 

\item [Anchoring of edges] - all fitting  polynomials have to be correctly
anchored at the edges to work well. However in many cases the edge of the chip
has a different behaviour than the middle. Usually the  edges show  rapid
decline in intensity due to the vignetting of optics or due to lower detector quantum efficiency 
on low resolution spectra covering whole visible range both at UV and IR
regions. Such curve with flat middle and rapidly falling edges is very
difficult to fit with low order polynomials. 

\item [Flexibility of fitting curve] - The big problem of fitting curves is
connected with their flexibility. The curve should be rigid enough to bridge
consistently the continuum gap over strong  wide lines (Hydrogen, Helium) and
flexible enough the fit well the changes in curvature of continuum. This is
especially important for normalising spectra of hot  stars  (strong He and
Hydrogen Balmer lines) stars or rapidly rotating stars with broadened lines.
Too high orders of fit tend to enter into the line, low order do not copy the
tilt of continua at the edges well. 

Majority of legacy spectral tools use Chebyschev or Legendre polynomials and
cubic splines (e.g. IRAF {\tt \small splot}, MIDAS {\tt\small XAlice} or
Starlink {\tt\small DIPSO}).  What is needed is a fitting routine that mimics the
freehand drawing of the experienced astronomer. One of the  little  known but
very powerful procedure is INTEP \citep{1982PDAO...16...67H}, which  uses
Hermittean polynomials, It has been successfully used many years  for
normalising spectra of Be stars in SPEFO \citep{skoda:1996} and recently
implemented in SPLAT-VO. Special attention was given to freehand drawing
algorithms by  \citet{akima:1970}. His procedure is implemented in SPLAT-VO as
well.

\item [Range of data points included into fit]  Some programs (e.g. IRAF {\tt\small splot})
allow to  fit only the  existing data points but some regions (e.g. wide lines)
may be excluded by giving the number of distinct regions. The more flexible
approach (e.g. in SPEFO, SPLAT-VO) is the possibility of placing individual
control points everywhere (e.g. above the apparent continuum). The
extrapolation of fitting line beyond the available data range can help to decide about
the best parameters (polynomial order, number of spline segments)   from 
its global behaviour (smoothness of curve and  asymptotic approach
to the particular value). 

\item [The problem of physical continuum]   The continuum may be not seen on some
objects so it cannot be fitted using the available data points.  Problem  in
cooler stars is to locate the real position of the continuum - it may not be present at
all due to severe blend of lines or molecular bands - then a model is needed.

It is  extremely difficult to normalise certain stars with unusual profiles ---
like stars with rapidly expanding envelopes (post-AGB outbursts, nova shells),
having strong emission lines  with P-Cyg profile dominating whole spectrum.

For some stars with extremely strong emission lines (e.g. some Be and B[e]
stars)  the dynamic range of 16-bit ADC is insufficient to obtain the
unsaturated line together with a well exposed  continuum. Often the continuum is hidden in
background noise.

\item [Normalising echelle spectra] A really challenging problem (which had not
yet been solved fully) is the normalisation of echelle spectra - especially
with wide absorption lines or complicated profiles (P Cyg, Be stars). Some wide
lines may span several echelle orders and so the individual orders have to be
precisely unblazed and then merged. This problem is very complex, depending on
construction of individual spectrograph, the observation strategy etc. See
\citet{2003adass..12..415S}, \citet{skosle:2004}, \citet{2001A&A...369.1048P}
or \citet{2002A&A...383..227E}.

\item [$\chi^2$ fitting procedures] The more advanced way of normalising is the
least squares fitting based on some theoretical model. It is very popular in
radioastronomy and X-ray astronomy ( XSPEC ) rather for estimation of the
global SED fit parameters (usually only as combination of several simple power
law functions). Quite useful in optical spectroscopy is the usage of
appropriate synthetic spectrum to find the real continuum in data. The
implementation of such a procedure in VO tools could call the Theory VO (TVO)
servers to get correct models (e.g. Kurucz) and perhaps to improve the models
by recomputing on the GRID in a iterative way.  Unfortunately for many
interesting objects models (even spectral classification) are only
speculative.

\end{description}

\subsection{Simple visual comparison}

A lot of the information about the behaviour of astronomical objects can be
estimated just by a visual inspection of a spectrum (spectral type,
peculiarity, emission) or a time series of spectra (pulsations, binarity).  The
basic method is the overplotting of many spectra in the same units and scale.
It may be very efficient with VO-enabled tools obtaining the number of spectra
cached  immediately from VO spectral servers  through Simple Spectra Access
Protocol (SSAP).  There are many possibilities of plots: 

\begin{itemize} 

\item Overplotting spectra of different objects in the same region

\item Overplotting the same object in different wavelength ranges (from X or UV over IR to radio). If the spectra are 
flux calibrated, it is the construction of spectra energy distribution (SED).  

\item Comparison of observation with models (rotationally broadened theoretical spectra) 

\item Plot of different lines overplotted in radial velocity scale  may give he
information about physical structure (e.g. for study of kinetic behaviour of
different elements in expanding envelopes \citep{kipper:2004}).

\end{itemize}

\subsection{Study of line profile variations } 

The high resolution spectra with high SNR may reveal on some objects small
variations of the   profile of spectral lines.  The study of LPV requires many
(even hundreds) of spectra to be overplotted to see  the changes well.
Sometimes the animation of changes in individual spectral line is very
impressive.  The low SNR data should be removed not to spoil the image. To
identify them  some form of colour highlighting should be implemented together
with single-stroke delete command to remove it from plotting list.  Such a plot
is helpful in asteroseismology or for estimating changes in stellar  winds.

\subsection{Dynamical  spectrum } 

It is sometimes called the gray representation or trailed spectrum. The basic
idea is to find the small time-dependent deviations  of individual line
profiles from some average. 

\begin{figure} 
\centering 
\includegraphics[width=1.00\linewidth]{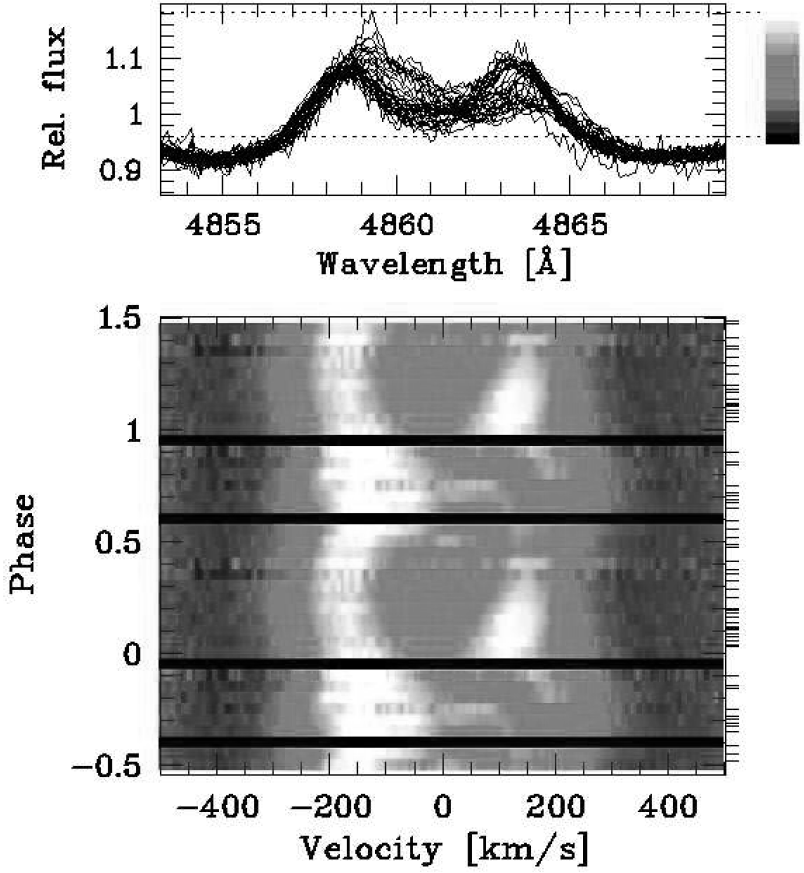}
\caption{Dynamical spectrum of H$\beta$  line profile variability of 59~Cyg.
Residuals from average profile of 38 spectra  are binned to 20 phase bins
corresponding to period 28.192 days. Two expanded cycles shown for clarity.
Individual profiles are overplotted above.  After \citet{maintz:2003}
\label{dynspec2}}
\end{figure}

\begin{figure} 
\centering 
\includegraphics[width=1.00\linewidth]{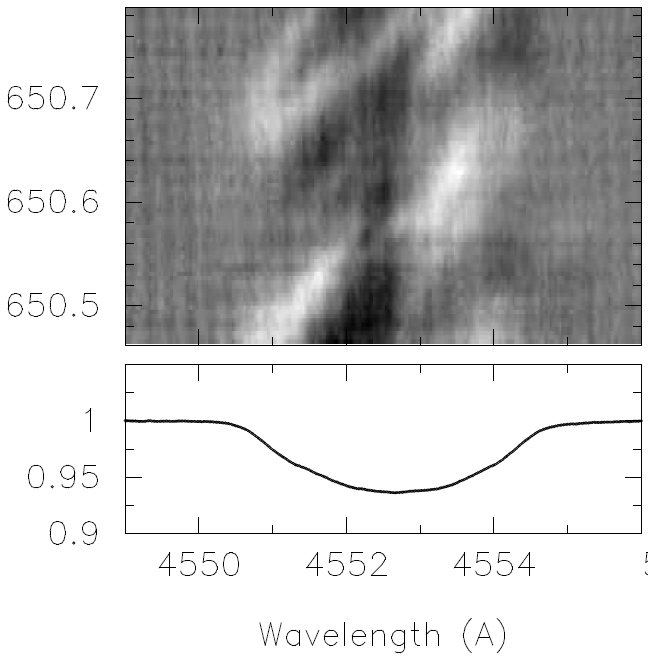}
\caption{Dynamical spectrum of SiIII  line profile variability of $\lambda$~Sco.
The average profile of 8 observing nights was used as a template shown below the panels. 
After \citet{uytterhoeven:2004}
\label{dynspec}}
\end{figure}

First the average of many high dispersion high SNR spectra (with removal of
outliers) is prepared (called template spectrum). Then each individual
spectrum in time series is either divided by the template (quotient spectrum)
or the template is subtracted from it (the differential spectrum). The group of
similar resulting intensities is given the same colour or  level of gray. 
See Fig.~\ref{dynspec2} and Fig.~\ref{dynspec}. More examples may be found e.g. 
in \citet{1999A&A...345..172D}, \citet{maintz:2003} or \cite{uytterhoeven:2004}.

The result is drawn in 2D image where on horizontal axis is a wavelength in the
line profile or corresponding RV relative to laboratory wavelength, on the
vertical axis the time of middle of observation (in HJD) or the circular phase
when the data are folded with certain period.  Valuable are
interactive features, like  zoom of whole dataset, removal of bad spectra from
series, adjustable colour cuts and look-up-tables etc.

\subsection{Measurement of radial velocity and higher moments of line profile}

The one of the important information received from spectrum is the radial
velocity of the object - e.g. the binarity can be revealed , or the possession
of extrasolar planet. RV is sometimes presented  as zero-th moment of the
spectral line profile. The higher moments of the line profile are important as
well. The first moment is equivalent width. The combination of higher moments
of line profile is a one of the possible ways of determination of non radial
pulsation modes numbers $l,m$ \citep{1992A&A...266..294A}.

\begin{description}

\item [RV by fitting profiles] This is a most common method implemented in
every spectral package.  The data points in given range of line profile are
directly fitted by Gauss, Lorentz or Voight function.  At least rough
normalisation is required. Problems occur in case of asymmetric and
complicated profiles (e.g. P-Cyg type).  

\item[RV estimation by  profile mirroring] 

It is a method used already  many years ago in oscilloscopic comparators. The
goal is to estimate the best match of the line profile and its flipped image
(mirror around y axis in certain wavelength). The mirrored  profile is being
shifted interactively until the best match is assessed.  This method (as
implemented in SPEFO) has been successfully used for measurement of RV of stars
with very complicated emission profiles (e.g. Be stars, luminous blue variables
with strong wind, nova outbursts).

The implementation should allow the region of interest (wing or core) in
vertical and the width of flipped area in horizontal direction to be easily
adjustable.  The reference line position is best taken from given  linelist
providing rest wavelength. The strength of this method lies in the capability
of measuring the match of core and  match of wings separately. See Fig.~\ref{mirrorwing}. 
More examples are given in \citet{2006IAUS..240E..48P}. This allows to study a different physics of
particular line-forming regions (e.g. shells, jets, stellar winds). 

\begin{figure} 
\centering 
\includegraphics[width=0.95\linewidth]{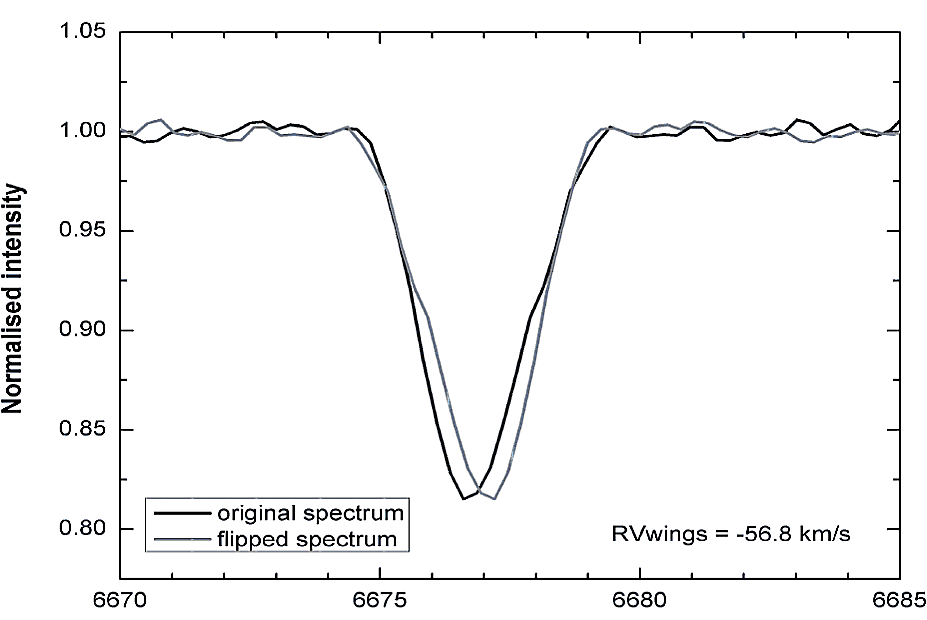}
\caption{The estimation of RV by mirroring of profile. This case shows the match of line wings.
\label{mirrorwing}}
\end{figure}

There is a close connection with bisector analysis, when the region of match
is controlled in different levels of depth of the line profile.  

\item[RV by cross correlation] It is often used to measure the RV of late type
stars with known spectral classification. It does not measure RV of individual
9iu
lines but a RV shift of the whole spectrum.  Many sharp lines are required and
a good template (usually a synthetic spectrum convolved to the similar
resolution and rotational broadening as observed one)
\citep{2000PASP..112..367B}.  Continuum normalisation is critical. 

Many legacy applications (IRAF package {\tt\small xrv} and {\tt\small RVSAO},
standalone program {\tt\small DAOSPEC}, MIDAS task {\tt\small XCORRELATE}
\ldots).  Often used for searching extrasolar planets (cross-correlation with
Iodine cell spectrum).  The cross-correlation is an ideal method for measurement of RV in echelle spectra (many
lines in different spectral regions may be combined together). 

\end{description}

\subsection{Measurement of equivalent width}

The measurement of equivalent width (EW) of a given line needs the
determination of area bordered by line profile and the continuum. This area is
transformed to the rectangular line with the depth of 1.0. The width of this
rectangle is called equivalent width. Emission lines by definition have
negative EW.

The measurement is sensitive to continuum placement. The extremely shallow lines give
large error in EW \citep{1997A&A...320..878V}.
EW are often used in abundance analysis. A good check of correct normalisation
is the comparison of EW of spectral lines of the same element in different
spectral regions \citep{2000A&A...358..553H}.

The comparison of EW  measured in spectrum of point-like solar
system bodies (minor planets) with high resolution solar spectrum from solar
telescopes gives the information about the intrinsic spectrograph properties
(stray light) and quality of pipeline reduction \citep{2002A&A...383..227E}.

\subsection{Bisector analysis}

\begin{figure} 
\centering 
\includegraphics[width=1.05\linewidth]{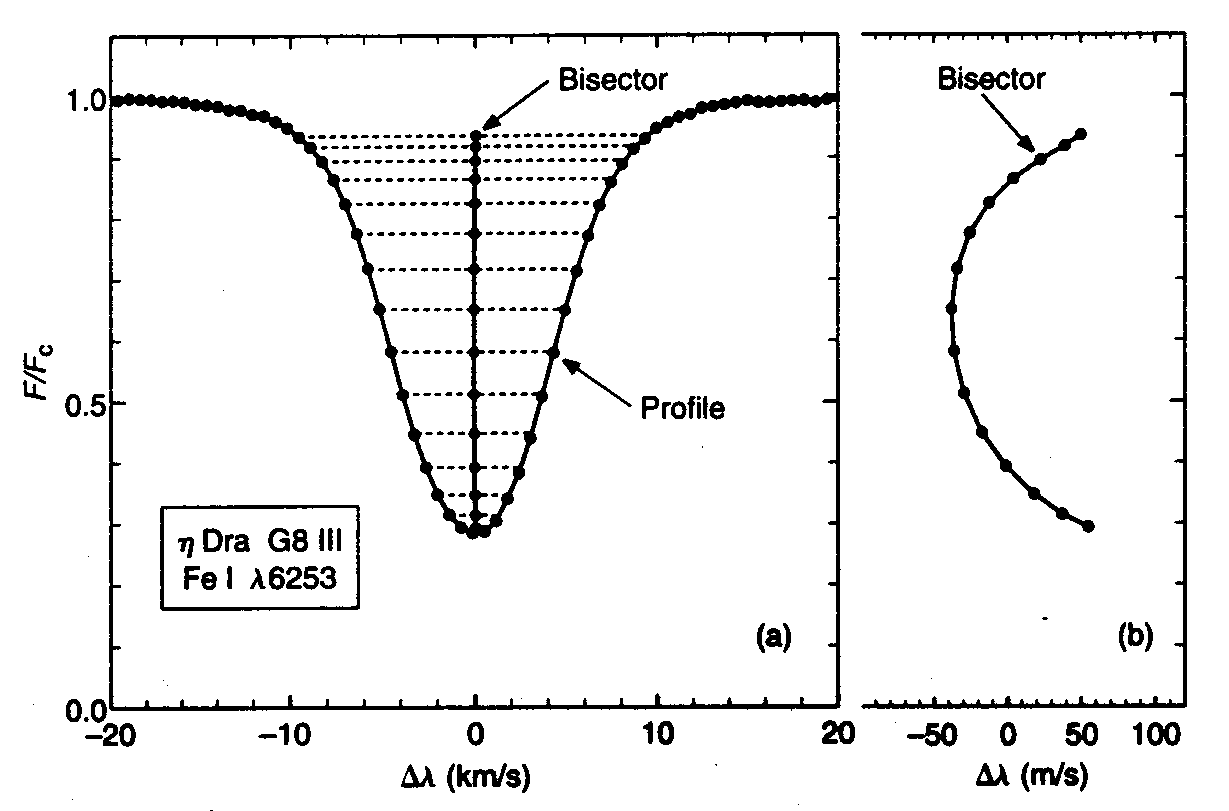}
\caption{The construction of bisector of line profile (left panel) and the zoomed bisector (right panel). 
From \citet{1982ApJ...255..200G}.
\label{bisector}}
\end{figure}

It is a method describing quantitatively the tiny asymmetry or subtle changes
in line profiles.  It is easily done by marking the middle of horizontal cuts
of profile in different line depths.  The line connecting such points (called bisector) is then
zoomed in horizontal (wavelength or RV) direction. See Fig.~\ref{bisector}.

The characteristic shape of  bisector gives the information about turbulence
fields (e.g. convection) in stellar photosphere, characterised by the value of
micro-turbulent velocity \citep{1982ApJ...255..200G, 2005PASP..117..711G} or
about other processes causing the tiny profile asymmetry.  It has been used
successfully for searching of extrasolar planets \citep{2001AJ....121.1136P} or
in asteroseismology.  Requires high resolution  (echelle) and high SNR
normalised spectra. Still some smoothing of edges in steeply declining line
wings is helpful \citep{2005A&A...442..775M}. 

\subsection{Rotational broadening}

Before comparing the synthetic spectrum with observed, the rotational
broadening has to be applied, especially for rapid rotators.  The estimate of
rotational velocity  $v \sin i$ is thus obtained.  

Although simple formula of convolution with parabolic kernel has been mostly
used, the precise computation is very complicated (the problem of limb
darkening) especially in non-LTE case \citep{2003ASPC..288..149H}.  

Sometimes a Fourier transformation of a line profile (with high SNR) is applied to get the first
estimate of rotational velocity \citep{2007ASPC..361..425G}.

\subsection{Period Analysis}

One of the most popular methods used in astronomy is the period analysis.  Its
aim is to find the hidden periods of variability of given object.  Sometimes
this period can be identified with some physical mechanism (e.g. orbital period of
binaries, rotational modulation or pulsations). Wide range of objects show the
multi-periodicity on various time scales (e.g. binary with pulsating components).
Ones the suspected period is found, the data may be folded accordingly, plotted
in circular phase corresponding this period. Very helpful is a interactive
capability of showing data folded while selecting different peaks at the
periodogram.

\begin{figure}[ht!] 
\centering
\includegraphics[width=1.00\linewidth]{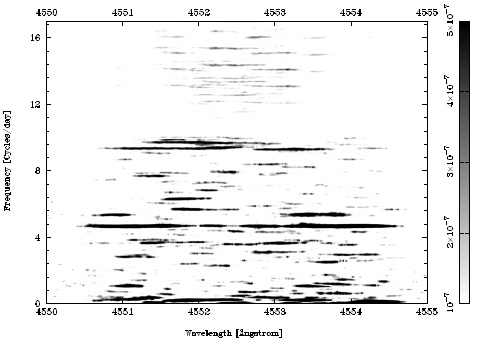}
\caption{Example of 2D periodogram of SiIII 4552~\AA\ line profile variability
of $\lambda$~Sco. 
After \citet{uytterhoeven:2004}.
\label{dynperiod}}
\end{figure}

The number of methods applied is large. but there are several very popular in astronomy:

\begin{itemize}

\item Folding techniques: Phase dispersion minimisation PDM
\citep{1978ApJ...224..953S}.

\item Fourier methods: Widely used algorithm CLEAN \citep{1987AJ.....93..968R}
or its modified  Lomb-Scargle version \citep{1976Ap&SS..39..447L,
1989ApJ...343..874S}.

\item Periodogram  of RV of bisector (sometimes called local RV). The periods
in position of bisectors in different line depths may give an idea where the
profile is affected by given non radial pulsation mode
\citep{2007ASPC..361..451K}.

\item Periodogram of line profile. It is a kind of gray representation with
period on vertical axis and the zoomed line profile on horizontal. See
Fig.~\ref{dynperiod}.  Works well for analysis of multi-periodical non-radial
pulsations \citep{1999A&A...345..172D}.

\end{itemize}

\section{Complex processing methods }

For the complex techniques given below, a lot of additional information is
required in addition to spectral data.  Programs require complicated
configuration files in given format and some interactive trials to find the
best results using the output from recent run as input to next one. They are
written often in FORTRAN without graphical interface or even the plotting
capabilities.  They are designed for batch runs driven by  parameter files.
 Their application is limited to special cases of certain
objects with appropriate physical properties.

\subsection{Doppler imaging }

It was introduced by \citet{1983PASP...95..565V} as a method allowing the
surface mapping of stellar spots.  First test were done on stars of RS CVn type
and on $\zeta$~Oph \citep{1983ApJ...275..661V}. Works well on rapid rotators
and needs a high resolution spectra with very high SNR (300--500). Tho whole
rotational period should be covered well, better several times.  When all the
requirements are met,  the map  of surface features (spots, nodes of non radial
pulsations) is obtained with very high accuracy.

\subsection {Doppler tomography}

It was introduced by \citet{1988MNRAS.235..269M}  for mapping the distribution
of emitting circumstellar matter in binary system. One of successful application
gave a picture of   accretion jets in Algols \citep{2004AN....325..229R}.  It
uses trailed spectrum  in velocity scale. The result is 2D image in velocity
space.  The transformation of radial velocity space to coordinate
space is ambiguous, which causes problems in interpretation of Doppler
tomograms. 

\subsection{Spectra disentangling}

This method allows to disentangle the spectra of individual stars in binary or
multiple systems even in case of heavy blending of lines.  It supposes the
changes in line profile are caused only by combination of Doppler shifted
components (no intrinsic variability of star).  The best solution of orbital
parameters and disentangled line profiles of individual stellar components are
found by least square global minimisation.  Automatically removes the telluric
lines with great precision.  Requires good orbital coverage and the estimate of
orbital parameters. Two approaches exist:  

\begin{description}

\item [Wavelength space disentangling] developed by \citet{1991ApJ...376..266B}
and improved by \citet{1994A&A...281..286S}.  It needs a large memory to store
sparse matrices, requires large computing power. Gives more reliable results

\item [Fourier space disentangling] introduced by \citet{1995A&AS..114..393H}
and \citet{1997A&AS..122..581H} in program KOREL. Another program available
today (still based on KOREL ideas) is  FDBINARY \citep{2004ASPC..318..111I}.
They  work in Fourier space, and transform the wavelengths into $\ln\,\lambda$.
They solve a small amount of linear equations, so they are   memory savvy and
can be run on even small computer.  The method, however,  requires  perfect
continuum fit (difficult to achieve at merged echelle spectra). On unreliably
rectified spectra strange artifacts may appear \citep{2004ASPC..318..111I}.

\end{description}

\section{VO-enabled tools}

Although there is a number of various tools written for the analysis of
spectra of astronomical objects (e.g IRAF, MIDAS, Starlink  packages {\tt
splot}, {\tt  spectool}, {\tt DIPSO}, {\tt XALICE} as well as numerous
single-purpose C, FORTRAN and IDL routines), the number of modern tools
understanding the VO protocols (especially spectra access protocol - SSAP)
is still very low.  We are giving a brief overview of their capabilities and
status of  development at the time of the submission of these proceedings (May 2007).

\subsection{VOSpec }

\begin{itemize}
\itemsep -1ex
\item Current version 2.5 
\item {\small http://esavo.esa.int/vospec/}
\item Developed at ESA
\item Very simple (mainly for building SED) 
\item Polynomial and Gaussian fits only
\item Blackbody fit 
\item No RV measurement built-in 
\item No complex operations with spectra allowed
\item It can be called directly from  VizieR
\item Can work with linelists through Simple Line Access Protocol (SLAP)
\item Theoretical VO (TVO) supported (synthetic spectra: Kurucz, disks)
\item Rapid development  
\item Integration of some methods in users own program possible
\item JAVA applet 
\item Contains PLASTIC VO-interoperability layer  
\item Dereddening of extragalactic objects built-in
\item Support of dimensional equations for units (DIMEQ, SCALEQ)
\item Rather complicated view of users data - needs to create SSA wrapper and
prepare VOTable even for one simple FITS spectrum before viewing 

\end{itemize}

\subsection{SpecView}

\begin{itemize}
\itemsep -1ex
\item Current version 2.13 
\item {\small http://www.stsci.edu/resources/software\_hardware/specview}
\item It is available as JAVA applet or  stand-alone application
\item Developed and supported  by STScI
\item Understands a number of formats from HST instruments and most NASA/ESA satellites
\item Understands general FITS in form of binary tables
\item Does not handle simple 1D FITS  (binary images with CDELT, CRVAL)
\item Easy work is with local data (if binary or ASCII tables)
\item Well  tailored for practical  spectral analysis
\item Powerful linelists (many included)
\item Fitting of models by  $\chi^2$ --- even user models may be constructed
\item Built-in number of standard stars spectra and models for various physical conditions
\item Whole Kurucz library of spectra available for immediate display.
\item Only simple polynomials for visual fit available
\item Dereddening, CLOUDY models built-in
\item Proper error propagation of fitted values 
\item Support of dimensional equations
\item Does not support PLASTIC
\end{itemize}

\subsection{SPLAT-VO}

\begin{itemize}
\itemsep -1ex
\item Current version 3.7 
\item {\small http://star-www.dur.ac.uk/~pdraper/splat/splat-vo/}
\item Supported by JCMT after closing Starlink
\item Most advanced for stellar astronomy (fitting by freehand drawing, INTEP)
\item RV measurement - both Gaussian fit and mirroring
\item Custom line list for individual spectra available
\item Wavelet analysis
\item Full featured data and editor and spreadsheet 
\item Publication quality output, powerful plotting options, annotations
\item Supported PLASTIC
\item Reads 1D FITS image files (e.g. rebinned IRAF multispec files) 
\item Asynchronous SSAP queries
\end{itemize}

\subsection{Period04}

\begin{itemize}
\itemsep -1ex
\item Current version 1.0.2 
\item {\small http://www.univie.ac.at/tops/Period04/}
\item Its predecessor was  widely used Period98
\item Period04 is rewritten in Java (and partly C) 
\item Needs formatted text files
\item VO-interface not supported, only local VOTable files
\item Several period finding methods built-in
\item Can handle even the regular period shifts
\end{itemize}

\subsection{FROG}

\begin{itemize}
\itemsep -1ex
\item Part of  Starlink Java package 
\item Included in last JCMT Starlink  release (Hokulei) 
\item {\small http://www.jach.hawaii.edu/software/starlink/}
\item VO-protocols built-in
\item Almost same capabilities as Period04
\item Unclear status of development after closing Starlink - all the links to documentation and files are now wrong 
\item Last version since 2004 - no further development noticed
\item The period searching engine available remotely as a SOAP service
\item Web service for Fourier transform
\item Collaborates with TOPCAT, but not PLASTIC
\item Easy but powerful for doing period analysis in VO environment
\end{itemize}

\section{Implementation strategies}

As was shown above, the current VO-enabled tools are already very flexible
to allow the large part of analysis of astronomical spectra to be directly
performed in the VO environment using the data through SSAP. However, many
``classical '' astronomers, working mostly with middle and high
resolution stellar spectra in visual wavelengths, will not use these tools
until most of the features described earlier are implemented. 

The functional behaviour of such implementations is very important as well.
To be quickly adopted by the conservative part of stellar  community, the
tool should be self-learning: it should allow interactive work with one or
couple of spectra  to adjust the parameters (ranges for measurement, degree
of fitting polynomials etc.).  But after tuning its behaviour it should be
able to apply the same recipes to large number of spectra with similar
characteristics. The very promising way how to do it is the concept of
workflows.

There is a important question where to put the engine for spectral analysis.
All tools today re the JAVA applications running as  VO-client on
astronomer's computer. Adding more features (new methods of data processing)
requires the new version of client to be downloaded and installed. Despite
the possible automatic installation through Java WebStart there is a serious
problem of size and complexity of such tool. Although the new functions
could be added as client plugins (similar to Firefox plugins), we do not
consider such a solution to be prospective  on a long-term scale as it is
against the original concept of VO. VO has been usually advertised  as a
facility allowing seamless federation of all distributed data archives and
computing resources providing data processing services through Web Services.

Certain amount of processing power should be thus left at the data providing
servers.  It may perform some simple operations  before sending the spectrum
to the client. For example  it may transform the  units (both wavelength and
flux),  cut out only the requested wavelengths (e.g. 20~A long section
centred around H$\alpha$) or even the server might perform the
normalisation of spectrum by approximate continuum fit - either computed
using given parameters of fit or just applying the fit included in
calibrated spectrum (some reduction programs produce  spectra in format
containing both the unrectified spectrum with intensity in ADU and the
continuum fit described by type and degree of polynomials and list of
control points).

But most of complicated processing (e.g. spectra disentangling, Doppler
tomography, period analysis,   should be rather performed on dedicated
problem-oriented servers using Web Services or similar technology. It is,
however,  difficult to convince the authors of powerful legacy (mostly
FORTRAN) applications to re-implement them according to requirements of VO
infrastructure (they do not know the JAVA, VO protocols etc.) and it is
dangerous to ask the software developer to blindly re-code such a
applications without detailed knowledge of the  employed mathematical
methods (including their bottlenecks and week points). 

Thus the only feasible way how to create most of such Web Services is to
create the VO-compatible wrappers calling the legacy application as  it was
designed using its native input and output file format. The collaboration
with authors is important anyway.

The utilisation of dedicated servers allows easy extension of processing
capabilities by adding another service to the same server (e.g. by adding
Apache-like server module) or by installing new dedicated server. The
natural way of solving complicated computing demanding problems is to deploy
such services on GRID.

There are still  tasks in astronomical spectra analysis requiring visual
interaction (e.g.  precise fitting of complicated continuum, measurement of RV
by profile mirroring or adjusting the colour scale of dynamical spectrum to
strengthen particular feature in it).  Such (usually simple) procedures may be
left on the client side. Today, when there is a very few applications written in
form of Web Services (to be deployed on server side), the extension of client
capabilities seems to be the easiest solution even for some server-oriented tasks 
(e.g. the cutout of selected spectral region).

\pagebreak 
\section{Conclusions}
Astronomical spectroscopy uses a wide range of techniques with different
level of complexity to achieve its final goal --- to  estimate the most
precise and reliable  information about celestial objects. The  large part
of spectroscopic analysis today  has been accomplished by several
independent non VO-compatible legacy packages, where  each works  with
different local files in its own data format. Analysis of large number of
spectra is thus very tedious work requiring good data bookkeeping. 

Accomplishing the analysis in VO infrastructure may benefit from automatic
aggregation of distributed archive resources (e.g. the multispectral
research), seamless on-the-fly data conversion, common interoperability of
all tools  (using PLASTIC protocol) and powerful graphical visualisation of
measured and derived quantities (e.g. in VOPlot or Mirage). 

By introduction of  modern  VO-aware tools into the  astronomical spectral
analysis  a remarkable increase of effectivity of astronomical research can
be achieved.

\section*{Acknowledgements}
This work has been supported by grant GACR 205/06/0584 and EURO-VO DCA WP6.
The Astronomical Institute Ond\v{r}ejov is supported by project AV0Z10030501

\end{document}